\documentclass[CEJP,DVI]{cej} 
\usepackage{layout}
\usepackage[OT4]{fontenc}
\usepackage{amsmath}
\usepackage{textcomp}
\usepackage{hyperref}

\title{Neutrinos from the Early Universe and Physics Beyond Standard Models}

\articletype{Review Article}

\author{Daniela~Kirilova\inst{1}\email{dani@astro.bas.bg}
        }

\institute{
     \inst{1} Institute of Astronomy and NAO, Bulgarian Academy of Sciences,\\
     Tsarigradsko Shosse, 72, Sofia, Bulgaria\\
          }

\abstract{Neutrino oscillations present the only robust example of
experimentally detected physics beyond the standard model.
 This review discusses the established and
several hypothetical beyond standard models neutrino characteristics
and their cosmological effects and constraints.
   Particularly, the contemporary cosmological constraints on the number of neutrino
   families, neutrino mass differences and mixing, lepton asymmetry
   in the neutrino sector, neutrino masses, light sterile neutrino
   are briefly reviewed.}

\keywords{neutrino \*\ beyond standard model \*\ cosmology \*\
 cosmological constraints \*\
 BBN}
 \pacs{14.60.Lm, 98.80.-k, 14.60.St.,26.35.+c,98.80.Cq}

\begin{document}
\maketitle


\section{Introduction}

 Physics Beyond the Standard Models (BSMs), i.e. beyond Electro Weak
Model and beyond Standard Cosmological Model (SCM, also called
$\lambda$ Cold Dark Matter model) is required for the explanation of
the astrophysical and cosmological observational data. Namely, the
contemporary SCM, contains considerable BSMs components - the so
called dark energy  (DE)  and dark matter (DM), both with yet
unrevealed nature, alas. These constitute  96\% of the universe
matter today, and play a considerable role at the matter dominated
epoch, i.e. at later stages of the Universe evolution!

 BSMs physics is needed also for revealing the nature and the
characteristics of the inflaton (the particle/field responsible for
inflationary expansion stage) and CP-violation (CPV) or/and
B-violation (BV) mechanisms. These are expected necessary
ingredients in the theories of inflation and baryon asymmetry
generation, which are the most widely accepted today hypotheses
providing natural explanations of numerous intriguing observational
characteristics of our universe.

The inflationary theory explains naturally and elegantly the initial
conditions of the universe in the pre-Friedmann epoch, namely: the
extraordinary homogeneity and isotropy at large scales of the
universe at its present epoch;  its unique isotropy at the Cosmic
Microwave Background (CMB) formation epoch (when the universe was
$\sim 380 000$ years old); its unique flatness and the pattern of
structures it has. Besides, the inflationary early stage explains
the lack of topological defects in the universe. While the baryon
asymmetry generation models explain the locally observed
matter-antimatter asymmetry of the universe.

     To understand these cosmological puzzles {\it physics  BSMs is required} either
to propose the candidates for DM, DE, inflation and baryon charge
carrying field,  or to change/extend the theoretical basis of SMs
(propose alternative gravitational theory, extended standard model
 or grand unified theory), etc. Alas, after many years of research
 there are no firm experimental detection of these BSM candidates, only
 experimental and observational constraints
 on the hypothetical candidates or/and theories exist.

 On the other hand,
 we have been already the lucky witnesses of the {\it experimental establishment of the BSM
 physics} in the neutrino sector.
Experimental data on neutrino oscillations  firmly determined three
neutrino mixing angles and three mass differences, corresponding to
the existence of at least two non-zero neutrino masses. The concrete
neutrino mass pattern and possible CPV mechanism
 are to be detected in near future.
Thus, the neutrino experimental data ruled out  the Standard Models
assumptions about zero neutrino masses and mixing and about flavor
lepton number (L) conservation.

   {\it Cosmology provides
complementary knowledge} about neutrino and BSM physics in the
neutrino sector, because  neutrino had a considerable influence on
the processes during different  epochs of the universe evolution. At
the hot early universe stage, radiation dominated (RD) stage, light
neutrinos were essential ingredients of the universe density,
determining the dynamics of the universe.~\footnote{ DE and DM had
negligible dynamical influence at RD stage}

Neutrinos  played also an essential role in different processes as
for example  Big Bang Nucleosynthesis (BBN). In particular, electron
neutrino participated in the pre-BBN neutron-proton transitions,
that took  place during the first seconds, and nucleons freezing,
and thus they influenced considerably  the primordial production of
the light elements  (BBN) during the first minutes of the universe.
Hence, BBN is very sensitive to the number  of the light neutrino
types, neutrino characteristics, neutrino chemical potentials, the
possible presence of sterile neutrino, etc. BBN is capable to
differentiate  different neutrino flavors, because $\nu_e$
participates into proton-neutron transitions in the pre-BBN epoch,
essential for yields of the primordially produced elements, while
$\nu_{\mu}$ and $\nu_{\tau}$ do not exert kinetic effect on BBN.

At later stages of the universe evolution ($T<$ eV) relic neutrinos,
contributing to the matter density, influenced CMB anisotropies,
~\footnote{ as far as at least one of the neutrino species became
non-relativistic (using the information about neutrino mass
differences from the neutrino oscillations data)} played a role in
the formation of galaxies and their structures.  CMB and LSS, being
sensitive to the total neutrino density and provide information
about the neutrino masses and number of neutrino
species.~\footnote{with present day accuracy the CMB and the LSS
data are flavor blind.}

Hence, although the relic neutrinos, called  Cosmic Neutrino
Background (CNB) are not yet directly detected, strong observational
evidence for CNB and stringent cosmological constraints on relic
neutrino characteristics exist from BBN, CMB and LSS data. In
particular, the determinations of light elements abundances and BBN
theory predictions are used to put stringent constraints on neutrino
characteristics (the effective number of relativistic particles,
lepton asymmetry, sterile neutrino characteristics, neutrino mass
differences and mixings). While CMB and LSS data provide constraints
on neutrino masses and neutrino number density  corresponding to CMB
and LSS formation epochs.

In summary: It is important to explore the cosmological constraints
on neutrino on one hand, because neutrinos  are messengers from the
very young  hot universe (CNB formation epoch corresponds to the
first seconds of universe when the universe plasma had $T \sim $
MeV) and thus they provide information about the physical conditions
od the early universe; on the other hand cosmology provides
 knowledge about the properties of neutrino, which is complimentary
 to the information coming from particle physics.

This review discusses the already established and also several
hypothetical
 BSMs neutrino characteristics
and their cosmological effects and constraints.
  Namely, we review the cosmological role of light neutrino, neutrino
  oscillations and possible lepton asymmetry in the neutrino sector.
  We present  contemporary cosmological constraints on neutrino properties,
   obtained on the basis of  astrophysical and cosmological data.
   Particularly, the cosmological constraints on the number of neutrino
   families, neutrino mass differences and mixing, lepton asymmetry hidden
   in the neutrino sector, neutrino masses, sterile neutrino possible
   characteristics, etc. are  reviewed.

  In the next section we discuss  SCM predictions concerning relic
  neutrino. In the third section  established BSMs neutrino physics
 and some hypothetical BSMs neutrino characteristics are discussed. Fourth and
 fifth sections review  BBN constraints on neutrino BSMs characteristics,
 in particular the constraints  on  sterile
 neutrino, active-sterile neutrino oscillations,
 neutrino-antineutrino asymmetry,  the dark radiation problem.

\section{Neutrino Predicted by SMs}
\subsection{Neutrino in Standard Electro Weak Model}
 In the Standard Electro Weak Model neutrinos are massless, spin 1/2 fermions with weak
interactions, i.e.  SU(2)$_W$ doublets. There exist 3 neutrino
flavors of light neutrinos, namely electron neutrino $\nu_e$, muon
neutrino $\nu_{\mu}$ and tau neutrino $\nu_{\tau}$. The  number of
light neutrino types (with masses $m< m_Z/2$)  was experimentally
measured by four LEP experiments to be:
 $N_{\nu}=2.984 \pm 0.008$.

However, there are robust experimental evidence for BSMs physics in
the neutrino sector: Neutrinos oscillate, i.e. their mass
eigenstates  do not coincide with their flavor eigenstates, there
exist neutrino mixing and non-zero neutrino mass differences.
Neutrino oscillations data proved that neutrinos are massive. The
origin of neutrino masses is still unknown.   Various mass
generation mechanisms, usually involving the introduction of extra
particles, are discussed.

 Besides, BSMs physics predicts other types of neutrinos,
not coupled to $Z$, called "sterile" $\nu_s$, i.e. SU(2)$_W$
singlets, not having the ordinary weak interactions. Sterile
neutrinos may be produced in GUT models,  in models with large extra
dimensions, Manyfold Universe models, mirror matter models, or by
neutrino oscillations, etc.  Cosmological effects of $\nu_s$ and
bounds on light $\nu_s$ will be discussed in more detail in the
following sections.

\subsection{Relic neutrino characteristics predicted in SCM}

According to the SCM at radiation dominated stage of the universe
neutrinos were dynamically important component, because  their
energy density was comparable to that of photons:
\begin{equation}
\rho_{\nu}=7/8(T_{\nu}/T)^4N_{eff}\rho_{\gamma}(T).
\end{equation}
where $N_{eff}$ is the effective number of the relativistic neutrino
species.  Hence, $\nu_e$,  $\nu_{\mu}$ and  $\nu_{\tau}$ influenced
considerably the expansion rate of the universe $H \sim \sqrt{8 \pi
G_N \rho/3}$.

At $T>1$ MeV neutrinos  were kept in equilibrium because their
interactions with other particles of the hot universe plasma, were
faster than the expansion rate of the universe: $\Gamma_W \sim
\sigma(E_{\nu}) n_{\nu}(T)>H$. Hence, neutrinos had an equilibrium
Fermi-Dirac (FD) energy distribution
\begin{equation}
n_{\nu}^{eq}=(1+\exp((E-\mu)/T))^{-1}.
\end{equation}
and their temperature coincided with that of electrons and photons.

As the Universe expanded and cooled $\Gamma$ decreased faster than
$H$, due to the decrease of energy and the dilution of particle
densities. Hence, roughly at $T \le 1$ MeV the interactions of
neutrino could no longer keep neutrino in equilibrium  and neutrinos
"froze out", i.e. neutrino species decoupled.  Since then neutrinos
propagated freely, i.e. cosmic neutrino background was formed. In
SCM neutrinos were assumed massless and, therefore, they kept their
FD distribution after decoupling. In the SCM the lepton asymmetry in
the neutrino sector was assumed zero.

The decoupling T slightly differed for different types of neutrino.
Namely $T^{dec}_{\nu_ e} \sim 2$ MeV, while $T^{dec}_{\nu_\mu, \tau}
\sim 3$ MeV.This is due to the fact that  electron neutrino, besides
neutral current interactions experienced also charged current
interactions (as far as electrons and positrons were still present
in the universe plasma at the time of neutrino decoupling, while
muons and tau leptons have already disappeared).  Neutrinos kept
their equilibrium  FD distribution after decoupling because of the
extreme smallness of neutrino mass.

At $T_e \sim m_e$ only photons were heated by $e^{+}e^{-}$ -
annihilation. Since then  neutrino temperature remained lower than
the temperature of the photons $T _{\gamma}$ by a factor
$(4/11)^{1/3}$.

 Actually, neutrinos shared  a small part of the entropy release
 because neutrino decoupling  epoch was close to $m_e$ and also because
 the decoupling was not instantaneous. The account for
non-instantaneous decoupling, for  QED finite temperature effects
and for flavor neutrino oscillations  slightly changes the predicted
neutrino particle and energy densities~\cite{j1,j2}.
 Thus, the expected number of neutrino species  is $N_{\nu} =
 3.046$ not 3, the neutrino number
 density is $n_{\nu}=339.3$ cm$^{-3}$ (per three neutrino species), not
 $335.7$ cm$^{-3}$. The temperature of CNB today is predicted to be  $T=1.9$
 K.

 Hence, CNB neutrinos  are expected to be the most numerous
particles after  CMB photons. The relic neutrinos today contribute
negligible part to the total density due to their low temperature:

$$\Omega_{\nu}(t_0)=2\times 7/8 \times (\pi^2/30)T^4_{\nu,0}/\rho_c\sim 10^{-5}.$$

  Thus, {\it
SCM predicts massless relic neutrinos, with equilibrium Fermi-Dirac
distribution, zero chemical potential and effective number of the
relativistic neutrino species} $N_{eff}=3.046$.

   Though
numerous, CNB neutrinos have not been directly detected yet, because
of  neutrinos weak interactions and extremely low energy of the CNB
neutrinos today. However, due to the cosmological effects of
neutrinos on processes which have left observable today relics, {\it
CNB has been detected indirectly} and numerous constraints on
 relic neutrino and on  BSM neutrino characteristics  have been
obtained. For more details see for example the review
papers~\cite{rev1,rev2,rev3,rev4,rev5}. The constraints on light
neutrinos will be discussed in the following sections.

\section{Neutrino Beyond Standard Models - established and hypothetical}

Here we discuss several examples of BSM neutrino physics: deviations
from FD distribution, neutrino oscillations, neutrino masses,
neutrino-antineutrino asymmetry.

\subsection{Deviations from  Fermi-Dirac distribution}

Different BSM  models  predict  deviations  from FD distribution of
the CNB neutrino. Among them are neutrino oscillations, models with
non-zero lepton asymmetry, models with additional decaying particles
into neutrinos or decays of heavy neutrinos.

 {\it Flavor neutrino oscillations} lead to a slight change of the FD
 distribution~\cite{j3,j4}.
 Flavor oscillations with parameters favored by the atmospheric and solar neutrino data
 establish an equilibrium between active neutrino species before BBN  epoch.
 Thus, the number density of one neutrino species with the account of flavor oscillations becomes
  113 cm$^{-3}$ instead 112$^{-3}$ (predicted by the  SCM).

The presence of non-zero {\it lepton asymmetry}  is  capable to
distort strongly the neutrino spectrum. However, the value of the
neutrino-antineutrino asymmetry is now strongly constrained by BBN
in all sectors due to the presence of flavor oscillations~\cite{j5}
(see section 5).

{\it Fast active-sterile neutrino oscillations}, which could have
been  effective before the epoch of neutrino decoupling, slightly
influence active neutrino distributions, because the active neutrino
states can be  refilled due to interactions with the plasma.

However, {\it active-sterile neutrino oscillations, proceeding
after}  $\nu_e$ {\it decoupling}, i.e. with oscillation parameters:

$$\delta m^2\sin^4(2\vartheta)\le 10^{-7} \rm eV^2$$

and provided that $\nu_s$ state was not in equilibrium before the
start of oscillations $\delta N_s=\rho_{\nu_s}/\rho_{\nu_e}<1$, may
strongly distort neutrino energy
spectrum and deplete neutrino number density~\cite{j6,j7,j8}.

Thus, relic neutrino may have strongly distorted non-equilibrium
spectrum today~\cite{j9}. 

Besides, resonant neutrino active-sterile oscillations may generate
neutrino-antineutrino asymmetry~\cite{j7,j8,j28}, which also may
lead to distortions of the FD neutrino distribution.

 Eventual {\it decays of
neutrino and into neutrinos} are another source for deviations from
the equilibrium neutrino spectrum.

Possible {\it violation of spin statistics} and the cosmological
constraints on it were also explored~\cite{j10,j11}.

\subsection{Neutrino Oscillations}

Solar neutrino problem and atmospheric neutrino anomaly were
resolved by the phenomenon of neutrino oscillations, confirmed by
the data of terrestrial neutrino oscillations experiments. The
dominant neutrino oscillation channels to solve the atmospheric and
solar neutrino problems
  have been proved to be flavor neutrino oscillations.

 Thus, it has been observationally and experimentally proved that neutrinos oscillate,
 i.e.  there exists flavor neutrino mixing in vacuum and neutrino mass eigenstates $\nu_j$
 do not coincide with the
flavor eigenstates $\nu_f$,

\begin{equation}
\nu_f=\sum_{i=1}^3 U_{fi}\nu_i, \hspace{0.5cm} \delta m_{ij}^2
=m^2_j-m^2_i \ne 0, ~~~(i \ne j)
\end{equation}

Namely,
$$
\delta m^2_{12} \sim 7.5 \times 10^{-5} \rm eV^2$$

$$|\delta m_{31}^2|~\sim 2.4 \times 10^{-3} \rm eV^2$$

and  $\sin^2 \theta_{12} \sim 0.3$,
 $\sin^2\theta_{23} \sim 0.39$,
 $\sin^2\theta_{13} \sim 0.024$.

This implies non-zero neutrino mass and mixing and non-conservation
of the individual lepton charges $L_f$, which are BSMs
characteristics for neutrino.

In general neutrino oscillations occur in medium (matter
oscillations) inside stars, planets, universe plasma, etc. For
example the data from solar neutrino detectors showed evidence for
matter effects in the solar $\nu_e$ transitions.
 Matter oscillations in the early universe plasma were studied as
 well~\cite{nr}.
The medium distinguishes between different neutrino types due to
different interactions, which lead to different average potentials
$V_f$ for different neutrino types:
\begin{equation}
V_f=Q \pm L
\end{equation}
 \noindent where $f=e, \mu, \tau$, $Q=-bET^4/(\delta m^2M_W^2)$ is the so called non-local term,
$L=-aET^3L^{\alpha}/(\delta m^2)$ is the local term and $L^{\alpha}$
is given through the fermion asymmetries of the plasma, $a$ and $b$
are positive constants different for the different neutrino types,
$-L$ corresponds to the neutrino and $+L$ to the antineutrino case.

Thus, due to the medium  the oscillation pattern may change: In
general the medium suppresses oscillations by decreasing their
amplitude, in comparison with the vacuum oscillations one. However,
 the mixing in matter may become maximal,
independently of the value of the vacuum mixing angle, i.e. enhanced
oscillation transfer may occur in medium when a resonant condition
holds.

For neutrino in equilibrium in  the early Universe, when working in
terms of mean neutrino energy is acceptable, the resonant condition
reads:
\begin{equation}
Q \mp L=\cos2\vartheta.
\end{equation}
For $Q=0$ this is the well known Mikheev-Smirnov-Wolfenstein
effect~\cite{j28p}.  For naturally small
 lepton asymmetry of the order of the baryon one 
at high temperature of the early Universe when $Q>L$  resonant
neutrino transfer is possible both for neutrino and antineutrino
ensembles for $\delta m^2 <0$.
 At low $T$, when $L > Q$,  resonance in the neutrino ensemble occurs
 at  $\delta m^2>0$ , while in antineutrino ensemble the resonant is possible for
$\delta m^2 <0$.

 In non-equilibrium case when the distortion in the spectrum
distribution of neutrino in the early universe is considerable  the
resonance condition and the description of neutrino propagation
become more complicated~\cite{j29,j30}. 

\subsection{Neutrino masses}

Two non-zero mass differences have been measured by neutrino
oscillation experiments,
 therefore at least 2 neutrino types have different and
 non-zero masses.
 Neutrinos with a definite mass can be Dirac fermions or
Majorana particles. However, whatever their nature, SMs cannot
accommodate neutrino masses without considerable extensions
(additional particles and/or extra interactions).

  From atmospheric neutrino oscillations data a lower mass limit
  follows:
 at least one type of neutrino has mass exceeding $0.048$ eV. For 3
massive neutrinos with mass $m$, the energy density today is
expected to be:

$$\Omega_{\nu}=3m/(93.14 h^2 \rm eV^2)$$.

 Hence, the neutrino oscillations data put lower limit on the
 energy  density of relic neutrino: $\Omega_{\nu}>0.003$.

  The pattern of
 neutrino mass is not known.
The following possibilities for the neutrino mass spectrum exist,
namely
 {\it normal hierarchy:} $m_1 <m_2 << m_3$
 and  {\it inverted hierarchy:} $m_3 << m_1 < m_2$.

The absolute neutrino masses have not been directly measured yet,
neutrinoless double beta decay and beta decay experiments set an
upper limit to the neutrino mass $m<2.05$ eV  at $95 \%$ C.L.

 Cosmology provides the most stringent constraints
 on the total neutrino mass.
Having very tiny masses,  flavor neutrino should not play an
important role as a dark matter candidate,
 because they are hot dark matter and predicted LSS in a Universe filled
 with hot DM is incompatible with observations.
 On this cosmological consideration the upper limit on the neutrino relative
  density and on the neutrino masses are
 obtained:

 $$\Omega_{\nu}<0.02, m<0.66 \rm eV $$

\subsection{Sterile neutrino and active-sterile neutrino oscillations}

Recent reviews on the phenomenology of sterile (right handed)
neutrino and the experimental, astrophysical and cosmological
constraints on them can be found in refs.~\cite{j13,j14}.

In case of presence of sterile neutrinos active-sterile neutrino
oscillations may also have place. Although neutrino anomalies are
well described in terms of flavor neutrino oscillations, it has been
pointed that sub-leading active-sterile neutrino oscillations may
provide a better fit~\cite{j12}.

In case active-sterile neutrino oscillations occur the following
beyond SM physics may be expected:

i) Active-sterile oscillations (effective before neutrino
 decoupling)
 may excite sterile neutrinos  into
 equilibrium ~\cite{j4,j15,j16},
i.e. 4  or more light neutrino families, instead of 3 may exist;

ii) Active-sterile oscillations (effective after flavor neutrino
decoupling) may lead to strong distortion of  the neutrino energy
 spectrum from its equilibrium FD form~\cite
 {j6,j7};

 This neutrino energy spectrum distortion, caused by neutrino oscillations, and its
dependence on the level of initial
population of the  sterile neutrino was discussed in detail also  in ref.\cite{j19}. 
 Depending on the concrete oscillation parameters, this may be very strong
 effect -  up to 6 times stronger than
 the dynamical effect of neutrino oscillations.

  iii) Active-sterile oscillations may change neutrino-antineutrino asymmetry of
 the medium  (suppress or enhance preexisting
 asymmetry)~\cite{j27p,j28,j7,j8};

iv) Sterile neutrinos, produced by neutrino oscillations, in the
$1-2$ KeV mass range may be viable candidates for warm
DM~\cite{j19p,j19ppp};

Such WDM models with sterile neutrinos are as compatible to
cosmological observational data as CDM ones~\cite{j19pp}.
 For a
recent review of the status of sterile neutrino as DM candidates,
see ref.~\cite{jDM}.

v) Sterile neutrinos are employed in leptogenesis models, where CP-
violating  decays of  $\nu_s$ produce the locally observed baryon
asymmetry of the Universe.

  Sterile neutrinos and  active-sterile neutrino oscillations may
play important role for neutrino involved processes during different
epochs of the Universe, from which observable relics have been
already found, like BBN epoch and CMB and  LSS formation epochs, or
from which observable relics are expected - like  CNB formation
epoch.

The most stringent limits on sterile neutrino and on active-sterile
oscillation
parameters have been obtained from BBN considerations  ~\cite{j20,j21,j22,j23,j24,j25}, and will be
discussed in detail the next sections.

\subsection{The interplay between neutrino-antineutrino asymmetry and  oscillations}

 There exists interplay between neutrino oscillations and the
 lepton asymmetry L
 of the medium in the early universe~\cite{j47}.
 On one hand, neutrino oscillations change
neutrino-antineutrino asymmetry of the medium,  on the other hand L
effects neutrino oscillations.

{\it Flavor neutrino oscillations} with the oscillation parameters
fixed from the experimental measurements are able to equalize
possible relic asymmetry $L_f$ in different flavors before
BBN epoch~\cite{j5}.

{\it Active-sterile neutrino oscillations}, depending on the
concrete values of the oscillation parameters and the
characteristics of the medium,  may  either
{\it suppress pre-existing asymmetry}~\cite{j15,j16} 
 or {\it enhance it} (in MSW resonant active-sterile
    oscillations).
 L enhancement in MSW resonant active-sterile neutrino oscillations was found
 possible both in collisions dominated  oscillations~\cite{j28}
      for   $\delta m^2 > 10^{-5}$ eV $^2$ and in the collisionless
      case~\cite{j7,j8} for    $\delta m^2 < 10^{-7}$ eV$^2$.

{\it Relic L effects active-sterile neutrino oscillations}.
Depending on the concrete values of the parameters, describing
oscillations and the medium
{\it L may suppress oscillations}~\cite{j28,j29} 
                   {\it or enhance them}~\cite{j29,j30}.  

In BBN with active-sterile  neutrino oscillations  spectrum
distortion and L generation lead to different nucleon kinetics, and
modified BBN element production. These cosmological effects and the
cosmological constraints on BSM neutrino physics will be discussed
in more detail in the following section.

\subsection{Neutrino-antineutrino asymmetry}

 Lepton asymmetry
of the Universe L, usually defined as
$L=(n_l-n_{\bar{l}}/n_{\gamma})$,  is not measured yet and may be
 considerably  bigger than the baryon asymmetry $\beta \sim
6.10^{-10}$, which has been already measured with great precision
from CMB anisotropy data and BBN light elements abundances data.
Considerable L might be contained in $\nu$ sector, hence future
detection of the CNB would provide the possibility for L direct
measurement.

Today L may be measured/constrained only indirectly: through its
effect on other processes, which have left observable traces in the
universe: light element abundances from BBN, CMB, LSS, etc.

Non-zero L has the following effect: it increases the radiation
energy density, hence it leads to faster universe expansion and
delays matter/radiation equality epoch. CMB, BBN  and LSS feel the
total neutrino density, thus they constrain L due to its dynamical
effect.

Besides this effect, BBN feels also {\it the kinetic effect of
non-zero electron neutrino asymmetry} exerted on the pre-BBN n-p
transitions. Thus, more restrictive BBN constraints on $L_e$  exist.

In case of electron-sterile oscillations there exists {\it indirect
kinetic effect of L due to L-neutrino oscillations interplay}: even
very small L $L<0.01$, which dynamical and kinetic effect can be
neglected,  lead to changes in electron neutrino number density,
energy distribution, and oscillations pattern, and hence, to changes
of n/p kinetics and BBN~\cite{j29,j30}.

The cosmological effects of L and the stringent cosmological
constraints on neutrino-antineutrino asymmetry are discussed in the
last  section.  Extra neutrino species, as well as cosmological
effects of BSMs neutrino and the BBN constraints on it are discussed
in the next section.

\section{BBN -  Early Universe Probe and Standard Model Physics Test}

Big Bang Nucleosynthesis is the most early and precision probe for
physical conditions in early Universe and it presents also the most
reliable test for new physics at BBN energies. In particular, it
constrains BSMs characteristics of neutrinos.

 According to standard BBN 4 light nuclides: deuterium D, helium
 isotopes He-3 and He-4 and
lithium Li-7 were produced in non-negligible quantities during the
hot stage of the Universe evolution, in the brief period between the
first seconds and the first few minutes. The corresponding energy
interval is   1 - 0.01 MeV. Then the conditions (particle densities
and energy) were appropriate for the cosmic nuclear reactor to work.
Later due to universe expansion the temperature and the particle
densities decreased and the production of  heavier elements was
hindered.~\footnote{The latter were produced much later in stars and
processes in cosmic rays.}

{\it Standard  BBN is theoretically well established and
observationally and experimentally supported.} The theoretical
inputs include: the neutron lifetime $\tau_n$, the gravitational
constant $G_N$, the baryon-to-photon number density
$\eta=n_B/n_{\gamma}$, the nuclear rates. $\tau_n$ and $G_N$ are
precisely measured: $\tau_n=885.7 \pm 0.8$ s and $G_N=6.7087 \pm
0.001 \times 10^{-39}$ GeV$^{-2}$. Precise data on nuclear processes
rates from laboratory experiments at low energy (10 KeV –- 1 MeV) is
available. Until recently  $\eta$  was considered as the only
parameter of  the standard BBN and was estimated  from the data on
the primordial abundances of the 4 elements, produced in
considerable quantities during BBN. Today
 precise determination of $\eta$ was made from the data of
CMB anisotropies (which corresponds, however, to the much later
epoch - the epoch of CMB formation). I.e. there exists independent
on BBN measurement of $\eta$.

{\it Besides, precise astrophysical data on the predicted by
standard BBN  abundances of the light elements exists.}
 Light elements D, He-3, He-4, Li-7 were measured in different pristine environments, i.e.
in systems least contaminated by stellar evolution.\footnote{
Primordial abundances cannot be observed directly, because chemical
evolution after BBN has changed the primordial yields.}

Namely, D is measured in high redshift, low metalicity H-rich clouds
absorbing light from background QSA. He is measured in clouds of
ionized H (H II regions) of the most metal-poor blue compact
galaxies. Li is measured in old Population II (metal-poor) stars in
the spheroid of our Galaxy, which have very small metalicity:
$Z<1/10 000 Z_{Sun}$.

Then, the observed abundances are analyzed, accounting for different
processes that might have changed their primordial values. It is
known that D is only destroyed after its BBN formation, He-4 is only
enriched in stars, while He-3 and Li-7 have more complicated
evolution being both destroyed and produced in different processes
after BBN (He-3 in stars, Li-7 - destroyed in stars, produced in CR
interactions). \footnote{ Thus, primordial He-4 for example is
obtained by regression to zero metalicity.}

In standard BBN primordial yields of the elements depend on the
baryon-to photon ratio $\eta$. The predictions of standard BBN for
$\eta$ value determined form CMB are in excellent agreement with
observational data of D, He-3, He-4. For Li-7  the consistency is
not so good the observational data is by  factor 3 lower than the
standard BBN predictions (the so called lithium problem).  Hence, D
and He-4 are used as reliable probes for BSMs physics, relevant at
BBN energies. (He-3 has a complicated post BBN evolution, therefore
it is not used as cosmological a probe.)

The primordial values, obtained from the analysis of  on
astrophysical data of these elements are~\cite{j31,j32}:
$$D/H=(2.53 \pm 0.05)\times 10^{-5}$$
$$Y_p=0,2565 \pm 0.006.$$

In general BBN depends on all known interactions and, thus,
constrains their BSMs modifications. The primordial yields, depend
strongly on nucleons freezing, which is determined
 by the competition between  weak rates $\Gamma_w$  of n-p transitions and the
expansion rate $H(N_{eff})$. Therefore, BBN  is used as a probe of
non-standard physics leading to changes in
 $H$,
  $\Gamma_w$, pre-BBN nucleon kinetics or BBN itself.  In particular, BBN
  probes i) any additional light i.e. $m< MeV$, relativistic during BBN,
  particles species (generations), $N_{eff}$  ii) nonstandard  interactions relevant at BBN epoch,
  iii) departures from equilibrium distributions of particle densities of nucleons
  and leptons (caused by neutrino oscillations, lepton asymmetry, inhomogeneous distribution of baryons, etc.)

  BBN produced He-4 is known to be the best speedometer 
   and is usually used to
  constrain additional radiation.  It is also the most exact leptometer at the RD stage~\cite{j33a}.

Because of the closeness of the BBN epoch and the epoch of formation
of the cosmic neutrino background, relic neutrino characteristics
are most strongly constrained by BBN. In the following the effects
of neutrino on BBN and several BBN cosmological constraints on BSMs
neutrino characteristics will be reviewed.

\subsection{Cosmological Effects and BBN Constraints on Neutrino}
Neutrinos of all flavors influence considerably $H$ during  pre-BBN
epoch and during BBN.
 Neutrino
BSMs characteristics, like extra neutrino  species, chemical
potentials or/and lepton asymmetries in the neutrino sector,
neutrino active-sterile oscillations bringing into equilibrium
additional species, all lead to the increase of $H$, i.e. exert {\it
dynamical effect on BBN}. Thus  BBN feels BSM physics  leading to
the  effective number of neutrinos bigger than the standard value
 3.046, i.e. extra neutrino species, neutrino non standard
 interactions, neutrino oscillations, nonequilibrium energy distribution,
 neutrino chemical potentials, neutrino asymmetry, etc.

Besides, electron neutrino and antineutrino have {\it direct kinetic
effect on BBN}, participating in the n-p transitions in the pre-BBN
epoch. Due to that, BSM physics in the electron neutrino sector,
like chemical potentials and/or neutrino-antineutrino asymmetry,
neutrino electron-sterile oscillations, electron neutrino decays or
decays into electron neutrinos, etc.,  have stronger effect on BBN
than the corresponding ones in the muon or tau neutrino sectors,
which lead only to the dynamical effects on BBN.
\subsubsection{Cosmological Constraints on $N_{eff}$}

 The effective number of relativistic species $N_{eff}$ is strongly
constrained by
cosmological considerations of the BBN produced He-4~\cite{j36}.

BBN cosmological constraints, using recent He-4
data~\cite{j32} read: 
  $N_{eff}=3.8^{+0.8}_{-0.7}$ at 95\% CL.

  Using
conservative approach an upper bound  $\delta N_{eff}<1$ at 95 \% CL
has been obtained in ref. \cite{j37}. 
Much stringent bound was achieved recently, based on BBN produced D
abundance and CMB data~\cite{j39}:

$$N_{eff} =3.53^{+0.66}_{-0.63}.$$

Recent BBN analysis~\cite{j40}, 
based on $Y=0.2565 \pm 0.006$ provide the constraint $\delta N_{eff}
= 0.66 \pm 0.46$, which is consistent with $\delta N_{eff} = 0$ at
95\% C.L. According to that analysis:
 $$N_{eff}
=3.71^{+0.47}_{-0.45}.$$

 $\delta N_{eff} \sim 1$ is favored, while
$\delta N_{eff} \sim 2$ is
 disfavored at more than 95\% C.L.

 Mind however, that the standard BBN bounds on $N_{eff}$ are tightened in the presence of active
 sterile oscillations, which lead to overproduction of
 He-4~\cite{j9}.

 Before Plank CMB measurements had larger errors than BBN in
determining $\delta N_{eff}$. The cosmological data based on WMAP7,
BAO and $H_0$ measurements provided
much looser constraints $2.7<N_{eff}<6.2$~\cite{j38}. 
  Besides, in contrast to BBN,  CMB data are not sensitive to neutrino
flavor content and cannot distinguish  the deviations of neutrino
distribution from equilibrium from dynamical
effects~\cite{j40p,j40pp}.

 Recent Planck measurements combined with other CMB data (Planck, WMAP and other
CMB high l expts) provide the following constraints ~\cite{j41}:
   $$N_{eff} =
3.361^{+0.68}_{-0.64} \rm  ~at ~95\%~ C.L.$$
Adding to the analysis H and BAO data (95\% Planck +WP+high
          l+H0+BAO), the errors are
 comparable to the BBN ones:
          $$N_{eff} = 3.52^{+0.48}_{-0.45}.$$
Again higher than the standard value $\delta N_{eff} >0$ is favored.
I.e. there are some indications both from BBN and CMB epochs for
extra radiation.~\footnote{Besides, the observed tension b/n H
direct measurements and CMB and BAO data is relieved if extra
radiation exists, the best fit being $N_{eff} = 3.37$ (for low l
power spectrum).}

 Assuming no entropy release between BBN and CMB epochs
simultaneous CMB+BBN constraints (68\% Planck +WP+high l+Y(Aver et
al.)) can be obtained, namely:
 $$N_{eff} = 3.41 \pm 0.3$$

 If CMB plus BBN D data are combined (68\% Planck +WP+high
  l+D(Pettini+Cooke)) the following stringent constrained holds:
  $$N_{eff} = 3.02 \pm 0.27.$$

  Besides, CMB data allow simultaneous constraints on extra radiation  and the total flavor neutrino mass or
  sterile neutrino mass. Namely the bounds in case of  {\it massless sterile neutrino} read:

$$N_{eff} = 3.29^{+0.67}_{-0.64} \rm  ~(95\% ~Planck +WP+high ~l)
~and $$ $$m<0.6 \rm ~~eV.$$

When BAO data are added, the bounds on extra radiation slightly
tighten, but then the bounds on the mass of the flavor neutrinos
tightens considerably: $m<0.28$ eV, which presents  the most
stringent bound today on the total flavor neutrino mass.

Accepting $m=0.6$ eV and allowing for {\it massive sterile neutrino}
(which presence was indicated by MiniBoone, reactor and Gallium
anomalies) the following constraints on extra radiation and sterile
neutrino mass have been obtained:

$$N_{eff}< 3.91 \rm ~(95\% ~~Planck +WP+high ~l)~~and$$ $$m_s<0.59 \rm ~
eV.$$

 The latter bounds are marginally compatible with fully thermalized sterile neutrino
 with sub-eV  mass $m_s<0.5$ eV, necessary to explain the oscillations
 anomalies, called "dark radiation". This puzzle and possible solutions to it are discussed in more detail
 below.

\subsubsection{Dark radiation and eV neutrino saga}

 {\it Different types of cosmological indications suggested
additional relativistic density}: BBN and especially the abundance
of He-4 ~\cite{j32,j42}, 
 preferred higher radiation density
than the standard cosmological model prediction, namely
$3<N_{eff}^{BBN}<5$. The CMB and LSS data, as well as their combined
analysis also pointed to excess
radiation.~\cite{j43,j44,j45,j45p}.

Besides, {\it different neutrino oscillations experiments (reactor,
LSND, MiniBooNe, Gallium)  data  with different neutrino sources and
different detector technology  presented indications for neutrino
oscillations with 1 or 2 sterile neutrinos with sub-eV
masses}~\cite{white}. Phenomenological studies and global fits,
taking into account all relevant data have been
provided~\cite{j48,j49,j50}. 
Recent analyses of neutrino data prefer 3+1 neutrino oscillation
models~\cite{j48}.%

 Non-standard BBN models
including extra neutrinos were discussed. Analysis of different type
of cosmological data provided during the last years showed that the
sub-eV neutrinos would have been brought into thermal equilibrium
before BBN~\cite{th1,th2,th3,th4}, and thus it is restricted by
standard BBN.  Besides,  sterile neutrinos in the sub-eV range would
produce unacceptably big amount of  hot dark
matter (inconsistent with LSS data)~\cite{j46,dodelson06}.

Thus, the presence of  two additional $\nu_s$ is in tension both
with BBN and with LSS requirements~\cite{Giusarma11,j39,j45p}.
Recent cosmological data show slight preference for extra
radiation and showed that  3+2 are difficult to realize in $\Lambda$CDM~\cite{j46}.

Thus, CMB, LSS and BBN disfavored two additional thermalized extra
neutrino. As discussed in the previous subsection, cosmology favors
one additional sterile neutrino, but prefers it to
be lighter than eV~\cite{Giusarma11,j41}. 
 Recent  Planck results, however,  showed no convincing evidence for
extra relativistic neutrino species.

Deviations from the minimal cosmological model have been discussed.
To relax the cosmological constraints on DR modifications of
$\Lambda$CDM
 including additional radiation, change in matter density,
lepton asymmetry in the electron $\nu$ sector, etc. have been
explored.

L dynamical and direct kinetic effects have been considered
 as an explanation of the excess radiation. It was shown that
excess radiation cannot be explained by degenerate BBN~\cite{j53}.
However, in case its value is large  enough to
 suppress active-sterile oscillations the presence of  L  may  prevent
complete thermalization of the sterile neutrinos and help to
circumvent BBN and LSS constraints.

      It has been shown that additional sub-eV sterile neutrino
       may be allowed by BBN with   $L \ge
       0.08$ ~\cite{j47}. 
       Lower values $|L|>10^{-2}$ were obtained in
       ref.\cite{mirizzi}.
 The difference might be due
 to different approximations used.

 Other possibilities of DR problem solution, including modified BBN with decays of heavy neutrinos, were discussed as well~\cite{j51,j52}.

\subsection{Cosmological Effects and Constraints of Neutrino
Oscillations}

{\it Flavor oscillations have no considerable cosmological effect},
because of the close decoupling temperature, and hence almost equal
population of the different neutrino flavors. However, in case of
non-zero neutrino asymmetries, flavor oscillations lead to
redistribution and equilibration of the asymmetries in the different
sectors~\cite{j5,j53}. 
Hence,  the  restrictive BBN constraints to the asymmetry in the
electron neutrino sector applies to the other sectors as well.

Hypothetical neutrino-antineutrino mass differences may reproduce
baryon asymmetry of the universe ~\cite{bar}. 
There exist CPT violation scenarios generating difference between
neutrino and antineutrino populations. Via subsequent sphaleron
processes or B-L conserving GUT symmetries the observable baryon
asymmetry can be produced.

Fast {\it active-sterile neutrino oscillation}, effective before
neutrino decoupling,  increase the expansion rate $H$ by introducing
additional relativistic particle - $\nu_s$, i.e. they exert {\it
dynamical effect}. First  idea about the dynamical effect of extra
relativistic species belongs to Shvartsman (1969).
  This effect caused by oscillations was precisely studied in numerous
publications, starting with the pioneer one~\cite{j3}, where the
vacuum oscillation case was considered, and ref.~\cite{j15}, where
matter  neutrino oscillations were considered.

The dynamical effect of oscillations have been explored in numerous
publications, see for example the pioneer papers~\cite{j3,j15,j16}.
He-4 mass fraction serves as the best speedometer at BBN epoch.


Neutrino oscillations have also a{\it direct kinetic effect on BBN}
processes. Namely {\it oscillations influence electron neutrino and
antineutrino number densities and/or spectrum}, thus effecting the
weak rates of n-p transitions $\Gamma_w \sim G_F E_{\nu} n_{\nu_e}$.
This direct kinetic effect of oscillations may be due to

i) the {\it change of the particle densities of electron neutrino
and antineutrino} by fast electron-sterile neutrino oscillations
~\cite{j15,j16}; 

 ii) the {\it distortion of the energy spectrum distribution of electron
neutrino} caused by late electron-sterile neutrino oscillations,
proceeding after decoupling~\cite{j6,j7,j8,j19}; 

 iii) the  {\it change of the neutrino-antineutrino asymmetry}:
production of a considerable asymmetry ($L>0.01$ (capable to
influence n-p kinetics) in the electron neutrino sector by fast
resonant electron-sterile oscillations~\cite{j28,j7,j8}, 
  or suppression of a preexisting L.

 Late active-sterile oscillations, effective after decoupling of $\nu_f$, may
 have also {\it indirect kinetic effect on BBN} due to generation
 of small asymmetry in the electron neutrino sector~\cite{j8,j21} (as already discussed in previous sections).
 Although in this case the generated $L$  is too small to have
direct influence on BBN kinetics, {\it oscillations produced L
influence the oscillations pattern - suppresses or enhances
oscillations, which reflects in change of BBN}.~\footnote{Tiny relic
asymmetry, $L<< 0.01$, may also influence indirectly the kinetics of
BBN due to the neutrino oscillations-asymmetry
interplay~\cite{j29,j29a,j30}.}

Due to the different effects of neutrino oscillations on BBN, it is
a sensitive probe of neutrino oscillation parameters and of neutrino
characteristics.

Most precise constraints on the neutrino active-sterile oscillation
parameters in case of fast oscillations have been obtained in
ref~\cite{j22}. 

The case of electron-sterile oscillations, effective after neutrino
decoupling,  was considered in
refs.~\cite{j8,j20,j21,j23,j24}. 
The precise account of the energy spectrum distribution of
oscillating  neutrino allowed to extend the BBN constraints towards
very small mass differences - down to $10^{-9}$. The constraints
relax at small mixings due to the exact simultaneous account of the
asymmetry generated in oscillations. The change of the BBN
constraints due to initially non-zero sterile neutrino state was
considered as well.

\section{Lepton asymmetry - Cosmological Effects and Constraints}

\subsection{Cosmological Effects of L}
Several well studied cosmological effects of L are known:

 {\it The dynamical effect of L}:  It consists in the increase of
the radiation energy density due to non-zero L:

 $$\Delta N_{eff}=15/7[(\xi/\pi)^4+2(\xi/\pi)^2],$$
  where $\xi=\mu/T$ is the
 $\nu$ degeneracy parameter. This effect
leads to faster expansion, delaying matter/radiation equality epoch,
thus influencing BBN, CMB, evolution of perturbations and LSS.

 {\it The kinetic effect of L}: It is noticeable for big enough asymmetry in the
$\nu_e$ sector$|L_e|>0.01$. Then the different number densities of
$\nu_e$ and $\bar{\nu_e}$, lead to changes neutron-proton transfers
in the pre-BBN epoch:

$$\nu_e + n \leftrightarrow p + e^-$$
$$\bar{\nu_e} + p \leftrightarrow n + e^+$$
$$n \rightarrow p + e^- + \bar{\nu_e} $$

and, correspondingly, change the  BBN yields. (See for example
ref.~\cite{Simha&Steigman} and refs there in.)

L with much smaller values, $L<<0.01$ may influence BBN through its
interplay with neutrino oscillations, {\it indirect kinetic effect}.

\subsection{Cosmological constraints on L}

      BBN provides the most stringent constraints  on L.
      In BBN with the known flavor neutrino oscillations, degeneracies
      in different neutrino sectors equilibrate before BBN
due to flavor neutrino oscillations. Hence, the more stringent BBN
constraint on $L_e$ is distributed to other neutrino
types~\cite{j5,j54}. 
 Thus BBN provides the stringent cosmological bound:
  $$|L|<0.1.$$

  See also recent analysis~\cite{j40}.

 CMB and LSS provide  looser bounds~\cite{j55}. 
 Future CMB experiments will be more sensitive to neutrino asymmetry and provide
 comparable or better limits on
 L than the BBN ones.

In the case of modified BBN with active-sterile neutrino
oscillations, due to the interplay between L and these oscillations
very stringent constraints  on L can be obtained~\cite{j29,j30,j57}.
 BBN with late electron-sterile neutrino
oscillations may be the most precise leptometer: it provides the
possibility to measure and/or constrain L with values close to
baryon asymmetry value, \cite{j30}.

Inn case of  BBN with non-zero L  the limits on active-sterile
oscillation parameters are changed.  Namely, due to the capability
of L to suppress or enhance oscillations, it may eliminate, relax or
strengthen BBN constraints on neutrino oscillations.

It has been found that big enough L  inhibits oscillations, and thus
in case active-sterile neutrino oscillations are detected, this
relation between L and oscillations parameters may be interpreted as
an upper bound on $L$~\cite{j30,j47}:

$$L<(\delta m^2(eV^2))^{2/3}.$$

 For example, taking the indications for
active-sterile oscillations with $\delta m^2 \sim
10^{-5}$~\cite{HolandaSmirnov} the following estimate of the upper
limit on L  follows: $L<10^{-3.3}$.

\section{Conclusions}

Cosmology provides a powerful test for  BSMs neutrino
characteristics.  In
   particular,  cosmological constraints on the number of neutrino
   families, neutrino mass
   differences and mixing, lepton asymmetry hidden
   in the neutrino sector, neutrino masses, sterile neutrino possible
   characteristics, etc. exist.
   These contemporary cosmological bounds on light neutrino
properties, obtained on the basis of  astrophysical and cosmological
data, are much stronger than the experimentally available ones and
hence, provide precious information about neutrino.

\section*{Acknowledgements} The author  thanks the anonymous
referees for the comments and suggestions that helped to improve the
paper.  The author acknowledges the financial support for
participation into SEENET-MTP Workshop: Beyond the Standard Models -
BW2013, Vrnjacka Banja, Serbia, where this paper was initiated.


\begin{thebibliography}{99}

\bibitem{j1} Dolgov, A., Hansen S., Semikoz, D., {\em Nucl. Phys. B}
{\bf 503}, 426 (1997)
\bibitem{j2}Mangano, G., Miele, G., Pastor, S., Pinto, T.,
Pisanti, O., Serpico, P., {\em Nucl. Phys. B} {\bf 729}, 221 (2005)
\bibitem{rev1} A. Dolgov, {\em Phys.Rept.} {\bf 370} 333 (2002)
\bibitem{rev2} S. Hannestad, {\em Ann.Rev.Nucl.Part.Sci.} {\bf 56} 137
(2006);{\em Prog.Part.Nucl.Phys.} {\bf 65} 185 (2010)
\bibitem{rev3}  R. Lopez, S. Dodelson, A. Heckler, M. Turner
{\em Phys.Rev.Lett.} {\bf 82}  3952 (1999)
\bibitem{rev4} D. Kirilova, J.M. Frere, {\em New Astron.Rev.} {\bf 56}
169 (2012)
\bibitem{rev5} J. Lesgourgues, S. Pastor, {\em Adv.High
Energy Phys.} {\bf 2012} 608515 (2012)

\bibitem{j3}Dolgov, A., {\em Sov. J. Nucl. Phys.} {\bf 33}, (1981)
700.
\bibitem {j4} G. Mangano, G. Miele, S. Pastor, T. Pinto, O. Pisanti, P.
Serpico,  {\em  Nucl.Phys. B} 729, 221-234 (2005)
\bibitem{j5} A. Dolgov, S. Hansen, S. Pastor, S.Petcov, G.Raffelt, D.Semikoz,
2002, {\em Nucl. Phys. B} {\bf 632}, 363; Y.Y.Y. Wong, {\it Phys.
Rev. D}{\bf 66},  025015 (2002)
\bibitem{j6}Kirilova, D.,  {\em JINR preprint E2-88-301}, (1988)
\bibitem{j7} Kirilova, D., Chizhov, M.,  {\em in Neutrino96}, 478
(1996)
\bibitem{j8} Kirilova, D., Chizhov, M., {\em Phys. Lett. B} {\bf 393}, 375 (1997).
\bibitem{j9} Kirilova D.,  {\em Prog. Part. Nucl. Phys.} (2010)
\bibitem{j10} A.Dolgov, A. Smirnov, Phys.Lett. B621,  1-10  (2005)
\bibitem{j11} A. Barabash, A. Dolgov, R. Dvornicky, F. Simkovic, A.Yu. Smirnov, {\em Nucl.Phys. B} 783, 90-111 (2007)
\bibitem{j12} P. Holanda, A. Smirnov {\it Phys.Rev.D} {\bf 83} 113011
(2011)
\bibitem{j13} M. Drewers,  {\em Int.J.Mod.Phys.} {\bf E22}  1330019 (2013)
\bibitem{j14} Kusenko A., {\em Physics Reports}, {\bf 481},1 (2009)
\bibitem{j15}Barbieri, R., Dolgov, A., {\em Phys. Lett. B} {\bf 237},
440 (1990).
\bibitem{j16}Enqvist K., Kainulainen K., Thompson M.,  {\em Nucl.
Phys. B} {\bf 373}, 498 (1992).
\bibitem{j19}Kirilova, D., {\em Int. J. Mod. Phys.} {\bf D 13}, 831 (2004).

\bibitem{j19p} T.Asaka, M. Laine, M. Shaposhnikov, {\em JNEP} 01,091 (2007)
\bibitem{j19pp} A. Boyarsky, J. Lesgourgues, O Ruchayskij, M.Viel, {\em JCAP} 0905, 012 (2009).
\bibitem{j19ppp} K. Nakayama, F.Takahashi, T. T. Yanagida, arXiv:1405.4670 (2014)
\bibitem{j20}Kirilova, D., Chizhov, M., {\em Phys. Rev. D} {\bf 58},
073004 (1998)
\bibitem{jDM} A. Boyarsky, D. Iakubovskyi, O. Ruchayskiy, Physics of the Dark Universe, V.1, Issue 1, 136-154 (2012)
\bibitem{j21}Kirilova, D., Chizhov, M., {\em Nucl. Phys. B} {\bf 591}, 457 (2000).
\bibitem{j22}Dolgov  A., Villante F., {\em Nucl. Phys. B} {\bf 679},
  261 (2004).
\bibitem{j23}Kirilova D., {\em Int. J. Mod. Phys.} {\bf D 16}, 1197,
(2007)
\bibitem{j24}Kirilova D., Panayotova M., {\em JCAP}, {\bf 12}, 014 (2006)
 \bibitem{j25}Panayotova M., {\em  Bulg.J.Phys.} {\bf 38},
341 (2011)
\bibitem{j27p}R. Foot, R. R. Volkas {\it Phys.Rev.Lett.}{\bf 75} 4350
(1995); {\it Phys.Rev.D}{\bf 55} 5147 (1997); {\it Phys.Rev.D}{\bf
56} 6653 (1997);{\it Phys.Rev.D}{\bf 59} 029901 (1999)
\bibitem{j28}Foot R., Thomson, M., Volkas R. 1996: {\em Phys. Rev. D} {\bf 53},
R5349; X. Shi,  {\em Phys. Rev. D}  {\bf 54} 2753 (1996)

\bibitem{j47}Kirilova D., {\em Hyperfine Interact.} {\bf 215}  1-3, 111-118, (2013);
 {\em Proc. 5th International Symposium on Symmetries in Subatomic
Physics} 18-22 June, 2012, Groningen (2012).


\bibitem{j28p}S. Mikheyev, A. Smirnov, {\it Sov. J. Nucl.
Phys.} {\bf 42}, 913 (1985); {\it Nuovo Cimento}{\bf 9C}, 17 (1986);
L. Wolfenstein, {\it Phys. Rev.} {\bf D17}, 2369 (1978).

\bibitem{j29} Kirilova, D., Chizhov M., {\em Nucl. Phys. B} {\bf 534},447 (1998)
\bibitem{j29a} Kirilova, D., Chizhov M., 2000 {\em in Verbier 2000, Cosmology and particle physics} 433,
astro-ph/0101083.
\bibitem{j30} Kirilova D.,  {\em JCAP }{\bf 06}, 007 (2012)
\bibitem{nr}  D. Notzold, G. Raffelt, {\it Nucl. Phys.} {\bf B307}, 924 (1988)
\bibitem{j31} M. Pettini, R. Cooke,  {\it Mon.Not.Roy.Astron.Soc.}  {\bf 425} )
2477 (2012)
\bibitem{j32} Yu. I. Izotov, T. X. Thuan, {\em ApJ} {\bf 710}  L67 (2010)
\bibitem{j33a} Kirilova D., {\em BAJ} {\bf 15} 1-16 (2011)
\bibitem{j34} Steigman, G., {\em JCAP} {\bf04}  029 (2010)
\bibitem{j36}Iocco, F., Mangano, G., Miele, G., Pisanti, O., Serpico, P.
  D., {\em Phys. Rept.} {\bf 472}, 1(2009)
\bibitem{j37} Mangano G.,Serpico P., {\em Phys.Lett.B} {\bf 701} 296 (2011)
\bibitem{j38}E. Komatsu et al.(WMAP 7 year) {\em ApJ Suppl.} {\bf 192}  18 (2011)
\bibitem{j39}Nollet K., Holder G., arXiv: 1112.2683
\bibitem{j40} Steigman G.,  {\em  Adv.High Energy Phys.} {\bf 2012}  268321 (2012)
\bibitem{j40p} A. Cuocco, J. Lesgourgues, G. Mangano, S. Pastor, {\em PRD} 71, 123501 (2005)
\bibitem{j40pp} M. Shiraishi, K. Ichikava, K. Ichiki, N Sugiyama, M. Yannaguchi,  JCAP 0907, 005
(2009)
\bibitem{j41} Planck Collaboration, arXiv: 1303.5076

\bibitem{j42} Aver,E., Olive, K.A., Skillman, E.D., {\em JCAP} {\em 05}, 003 (2010a)
\bibitem{j43} Komatsu et al. 2011, arXiv:1001.4538 
\bibitem{j44} R. Keisler et al., {\it Ap.J} {\bf 743}, 28 (2011); arXiv:1105.3182
\bibitem{j45} Dunkley J. et al., {\it A.J} {\bf 739}, 52 (2011)
\bibitem{j45p}Hou Z. et al., {\em Phys.Rev. D} {\bf 87} 083008 (2013) 
\bibitem{j46} Hamman et al. 2012; Hamman J. et al., {\it JCAP} {\bf 1109}, 034 (2011); arXiv:1108.4136
J. Hamann, et. al. (2010) {\em Phys. Rev. Lett.}  {\bf 105} 181301.
\bibitem{white} K.N. Abazajian et al. Sterile Neutrinos: A White Paper e-Print:
arXiv:1204.5379 (2012)
\bibitem{th1} P. Di Bari, {\em PRD} 65, 043509 (2002)
\bibitem{th2} K. Abazajian, {\em Ap.Phys.} 19, 303 (2003)
\bibitem{th3} M. Cirelli, G. Marandella, A. Strumia, F. Vissani, {\em
NPB} 708, 215 (2005)
\bibitem{th4}S. Hannestad, I. Tamborra, T. Tram, {\em  JCAP} 1207, 025 (2012); arXiv:1204.5861

\bibitem{dodelson06} Dodelson S., Melchiorri A., Slosar A., {\it Phys. Rev. Lett.} {\bf
97}, 14130 (2006); astro/ph 0511500
\bibitem{Giusarma11} Giusarma E. et al, {\it Phys. Rev. D} {\bf 83} 115023 (2011)
\bibitem{mirizzi} Mirizzi A. et al.,  {\it Phys.Rev. D}  {\bf 86}053009 (2012) 
\bibitem{j48} C. Giunti, M. Laveder, {\em Phys. Rev. D} {\bf 84} 093006 (2011)
\bibitem{j49} Maltoni M., T. Schwetz, {\em Phys. Rev. D} {\bf 76}, 093005 (2007); J. Kopp, M. Maltoni,
T. Schwetz, {\em Phys.Rev.Lett.} {\bf  107}  091801 (2011)
\bibitem{j50} Karagiorgi G., {\em  Phys. Rev. D} {\bf 80}073001 (2009); arXiv:1110.3735 (2011)
\bibitem{j51}Dolgov A.,Kirilova D.,  {\it
Int.J.Mod.Phys.} {\bf A3}, 267 (1988); Kirilova D., School and
Workshop on Space Plasma Physics,83-89, AIP Conf. Proc., 1121,
31.08-07.09,Sozopol, Bulgaria, ed.I. Zhelyazkov, 2009
\bibitem{j52} O. Ruchayskiy, A. Ivashko, {\it JCAP} {\bf 1210} 014 (2012) 
\bibitem{j53} Mangano G., G. Miele, S. Pastor, O.
Pisanti, S. Sarikas, {\em  JCAP} {\bf 11}, 035 (2011);
\bibitem{bar} Barenboim G., Borissov L., Lykken J., Smirnov A., {\em JHEP}, 0210, 001 (2002)
\bibitem{j54} G. Mangano, G. Miele, S. Pastor, O. Pisanti, S. Sarikas, {\em
Phys.Lett. B} {\bf 708}  1 (2012)
\bibitem{j55}Lesgourgues, J., Pastor, S., {\em Physics Reports}, {\bf 429},
307 (2006).

\bibitem{j57} Kirilova D. {\em Prog. Part. Nucl. Phys.}{\bf 66}, 260 (2011)
\bibitem{Simha&Steigman} Simha, G. Steigman, {\it JCAP}{\bf 0808}
011 (2008)
\bibitem{HolandaSmirnov} P.C. de Holanda, A.Yu. Smirnov, {\em  Phys.Rev. D} 83,  113011 (2011)

\end{thebibliography}
\end{document}